\documentclass[epj]{svjour}
\usepackage{amsfonts}
\usepackage{amsmath}
\usepackage{amssymb}
\usepackage{epsfig}
\usepackage{graphicx}
\newcommand{\ar}{\arrowvert}

\newcommand{\be}{\begin{equation}}
\newcommand{\ee}{\end{equation}}
\newcommand{\ba}{\begin{eqnarray}}
\newcommand{\ea}{\end{eqnarray}}

\begin{document}
\title{On the violation of the holographic viscosity  versus entropy 
KSS bound in non relativistic systems}
\author{ Antonio Dobado, Felipe J. Llanes-Estrada}
 \institute{Departamento de F\'{\i}sica Te\'orica I, Universidad
Complutense de Madrid, 28040 Madrid, Spain}

\abstract{ A computation of the quotient of shear viscosity to
entropy density, or KSS number $\eta/s$ is performed, in the
non-relativistic and classical regime, first in Chiral Perturbation
Theory, and then in the $SO(g+1)/SO(g)$ Non-Linear Sigma Model in
the large $g$ limit. 
Both are field theories stemming from a renormalizable Sigma Model
but in spite of that, we explicitly calculate
how one undercomes the KSS bound by increasing the number of
degenerate pions sufficiently. However we argue that particle
production could still keep the validity of the KSS bound in the
weak sense. We also show how a large number of molecular
isomers (that we estimate in terms of simple molecular properties) could 
undercome the bound in the strong sense. This might be possible with 
carbon-based molecules. We finally argue that a measurement of $\eta/s$ 
in Heavy Ion Collisions might be turned into an upper bound on the number 
of hadron resonances.}

\authorrunning{Antonio Dobado and Felipe J. Llanes-Estrada}
\titlerunning{Violation of KSS bound in non-rel. systems}
\maketitle

\section{Introduction}

\begin{figure}
\psfig{figure=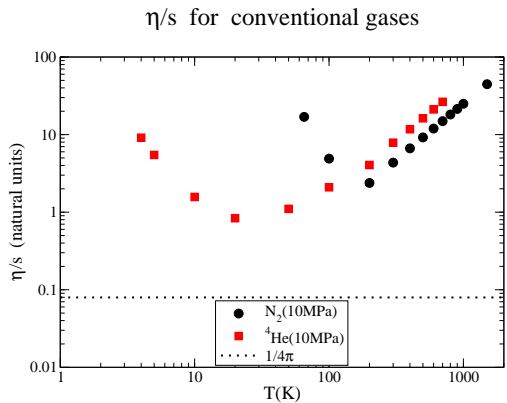,height=3.0in,angle=-90}
\caption{$\eta/s$ for molecular nitrogen and Helium gases together with
the bound of Kovtun, Son and Starinets. \vspace{-0.3cm}
\label{normalgases}}
\end{figure}

There is considerable interest in devising or reporting fluids
with the lowest possible value of the shear viscosity to entropy
density ratio, $\eta/s$. This ratio, dimensionless in natural
units, and taken at zero chemical potential, is responsible for
the damping of shear waves in a fluid with dispersion relation
$\omega+i k^2 (\eta/ s T)=0$. Few years ago, Kovtun, Son and
Starinets (KSS) \cite{Kovtun:2003wp} observed that in field
theories that have a gravity dual in higher dimension through the
holographic principle, the ratio could be estimated as a function
of the metric coefficients near a ``black brane''
$$
\frac{\eta}{s}= T f[g_{\alpha\beta}]
$$
and to their surprise, several feasible calculations with simple
metrics $g_{\alpha\beta}$ consistently yielded the value
$\eta/s=1/4\pi$. The field theories dual to these gravity
configurations are strongly coupled supersymmetric Yang-Mills
theories, far removed from our current physical picture of the
world. However, going to common substances whose viscosity and
entropy density values are tabulated, one finds (see fig.
\ref{normalgases})
that this ratio is at least an order of magnitude larger than
$1/4\pi$. In the spirit of other dimensionless numbers
characterizing fluid mechanics, such as the Reynolds or the
Prandtl number, we can define (now in arbitrary units) the $KSS$
number as \be KSS= \frac{k_B\eta }{\hbar s  } \ . \ee Given that
no fluid is known to have $KSS<1/(4\pi)$, the authors conjectured
that this value is a universal bound for the ratio (strong bound).
Since then, several works have reported on values for the ratio in
the range 0.08-0.3 for the quark and gluon liquid produced at RHIC
\cite{Gavin:2006xd}, 0.3 for cold, trapped atoms near a Feshbach
resonance \cite{Schafer:2007pr}, and theoretical work has shown
that the pion gas in the aftermath of a heavy-ion collision should
respect the bound \cite{Dobado:2006hw} (correcting earlier results from 
\cite{ChenNakano}), 
as also should the nuclear
matter formed in possible strange stars \cite{Laha:2007hs}.

As it was stressed in \cite{KSS} the KSS bound does not involve
the speed of light $c$ and hence is non trivial when applied to
non-relativistic systems.
It is not clear whether the bound is or is not appropriate for all such
systems.
The reason is that one can consider a gas with an increasing number of
different species of molecules so that the entropy density can be
made arbitrary large thus evading the bound. However it can be
argued this cannot happen in any respectable non-relativistic
system coming from a genuine relativistic quantum field theory
with a well defined ultraviolet (UV) completion (weak sense
bound). In this work we will show how the bound can be violated by
the non-relativistic classic limit of a $SO(g+1)/SO(g)$ Non-Linear
Sigma Model (NLSM) for a sufficient large $g$. As it is well known,
this Lagrangian density is the effective field theory of the
corresponding Linear Sigma Model (LSM), that is a renormalizable field 
theory (although the loophole of UV completion remains open, since the 
beta function is positive, one can call into question the relevance of 
this issue in the non-relativistic limit. 
Further discussion can be found in 
the recent paper  \cite{Cohen:2007qr}).

In order to expose the violation, we will
first consider the case of a non-relativistic classical pion gas
described by Chiral Perturbation Theory ($\chi$PT) and then we
will extend  the result to the $SO(g+1)/SO(g)$ Non-Linear Sigma
Model to see how the weak version of the bound can be avoided.
Then we will comment on a possible way out for the KSS conjecture
and finally we examine what one would need to achieve in molecular physics
to have a system  realized in Nature violating the bound.
Throughout the paper we distinguish ``KSS bound'' (meaning
$\eta/s>1/(4\pi)$ for an arbitrary system) from ``KSS conjecture'' (the
actual statement that this bound could be strict for relativistic field
theories in some more or less strong form depending on additional
assumptions).

\section{The bound applies for a non-relativistic pion system}
As it is well know the low-energy dynamics of pions can be
described by $\chi$PT \cite{ChPT}. At the lowest order this amount
to the NLSM based on the symmetry group $SU(2)_L\times SU(2)_R$
spontaneously broken to $SU(2)_{L+R}$. As a consequence the fields
have as domain the coset manifold $SU(2)_L\times
SU(2)_R/SU(2)_{L+R}=SO(4)/SO(3)=S^3$. This symmetry scheme is
manifest in the Lagrangian for the pion fields \be
\mathcal{L}_\chi=\frac{1}{2}g_{ab}\partial_{\mu}\pi^a\partial^{\mu}\pi^{b}+m^2f^2
\sqrt{1-\pi^2/f^2} \ee where the coset metrics is given by
$$g_{ab}=\delta_{ab}+\frac{\pi_a\pi_b}{f^2-\pi^2}\ ,$$ with $f\simeq 92
MeV$, $m \simeq 138 MeV$, $a,b$ running from $a,b=1$ to $a,b=g=3$
and $\pi^2=\pi^a\pi^a$. Note that we have added an explicit
symmetry breaking term proportional to $m^2$ to the NLSM
Lagrangian in order to take into account the pion mass $m$.

From the above Lagrangian it is possible to compute the  pion
elastic scattering amplitude which can be written as
$T_{abcd}=A(s,t,u)\delta_{ab}\delta_{cd}+...$ where we have not
shown explicitly the crossing terms. At low energies, which is the
relevant limit for the non-relativistic regime, the amplitude is
given by the Weinberg low energy theorem: $A=(s-m^2)/f^2$
\footnote{In fact the pion mass $m$ and the pion decay constant
$f$ appearing in this formula are modified by well known chiral
corrections but they are not relevant for our discussion here}.
The corresponding averaged cross section in the non-relativistic
limit is: \be \sigma=\frac{23m^2}{384\pi f^4} = \pi R^2 \ . \ee
 As this cross section is energy independent in this non-relativistic
limit we have introduced the effective pion radius $R$.

On the other hand the well known viscosity of a classical,
hard-sphere gas grows moderately for low temperature as a square
root, \be \label{hardvisco} \eta= \frac{5\sqrt{mT}}{16\sqrt{\pi}R^2}
\ee in terms of the hard sphere radius $R$ and mass $m$. Thus we can
use the above results to compute the viscosity of a non-relativistic
classical gas of pions which is given by:
\be \eta_\chi =
\frac{120\pi^{3/2}f^4}{23 m^{3/2}}\sqrt{T} \ . \ee

The entropy density of a Bose gas with $g$ different components is
conveniently taken as \be \label{Boseentropy}
s=\frac{g}{6\pi^2T^2} \int_9^\infty p^4dp\frac{E-\mu}{E}
\frac{e^{\beta (E-\mu)}}{[e^{\beta (E-\mu)}-1]^2} \ . \ee
 Note that we have introduced a chemical potential $\mu \le m$. Strictly
speaking this requires that the number of bosons (pions in our
case) is conserved by the interactions. However, from the $\chi$PT
Lagrangian we clearly see that interactions with any even number
of pions are allowed. In particular $2$ to $4$ pion reactions are
present whenever the center of mass energy is larger than the four
pion threshold located at $s=16m^2$. In spite of that, in the low
energy (non-relativistic) regime we are considering here, these
non pion number conserving processes are completely suppressed and
thus it is possible to introduce an effective chemical potential
$\mu$ associated to pion number. Consequently this parameter (or
equivalently the $n$ number density) can be arbitrary chosen
 for any temperature $T$
by assuming that the system is surrounded by a particle bath.

Now consider the cold $T<<m$, classical regime of the pion gas,
where the relativistic fugacity is taken small
 $z=e^{\beta(\mu-m)}<<1$. Defining as usual a thermal De Broglie
wavelength
$$
\lambda= \sqrt{\frac{2\pi}{mT}}
$$
the number density becomes \be n=\frac{gz}{\lambda^3} \ee and the
condition of classicity, or low  average occupation
number, reads $n<<g/\lambda^3$. Under this condition, eq.
(\ref{Boseentropy}) reduces to the formula of Sackur-Tetrode \be
s=n\left( \log \frac{g}{n\lambda^3}+\frac{5}{2} \right) \ee and
the KSS number becomes \be \label{KSSchiral} \frac{\eta_\chi}{s} =
\frac{240\sqrt{2}\pi^3}{23}\frac{f^4}{m^4}
\frac{m}{T}\frac{1}{n\lambda^3\left( \log
\frac{g}{n\lambda^3}+\frac{5}{2} \right)}\ . \ee

It is very easy to see that in the region of validity of this
formula, namely $T<<m$, $n \lambda^3 << g=3$ and $m \sim f$, we
have  $\eta_\chi/s \gg 1/ 4 \pi$. Therefore the KSS bound applies to
ChPT in the classical and non-relativistic limit. This model
corresponds to the low energy limit of Quantum Cromodynamics (QCD)
with $N_f=2$ and $N_c=3$.  Thus this version of QCD can be
considered the UV completion of the NLSM considered here.

One of the most remarkable features of the above formula is of
course the $g$ factor in the denominator. For physical pions in an
isospin triplet, this degeneracy is $g=3$. It would appear that
increasing the number of flavours in QCD, that implies increasing
the number of chiral Goldstone bosons (pions), eventually
undermines the KSS bound according to the above formula.
Nevertheless it is well known that one cannot increase the flavor
number arbitrarily in QCD without changing the derivative of the
$\beta$ function $\beta(g)= -g^3\left(11-2N_f/3 \right)/16\pi^2$
so that the quantum field theory is presumably not well defined.
In addition the computation that we have done here is based on the
NLSM based on the coset $SO(g+1)/SO(g)=S^g$. However the cosets
relevant for low energy QCD are $SU(N_f)_L\times SU(N_f)_R
/SU(N_f)_{L+R}$. Both families of cosets meet for $g=3$ and
$N_f=2$ but not in the general case. Thus our computation does not
describe low energy QCD for $g$ different from $3$.

\section{Violating the KSS bound in the large $g$ limit}

However the $SO(g+1)/SO(g)=S^g$ NLSM is the effective theory of  the
LSM based on the same groups. The LSM is a renormalizable Quantum
Field Theory (QFT) which is well defined perturbatively, although this 
may be not the case for some non-perturbative formulations because of the 
issue of triviality. It seems to be of help to the conjectured bound that 
the interaction decreases with the energy scale (the gas viscosity 
increases when the entropy also does) and maybe this is what will be left 
of the conjecture. This surely deserves further exploration.

The LSM Lagrangian is: 
\be
 \mathcal{L}=\frac{1}{2}\partial_\mu\Phi\partial^\mu 
\Phi^T-V(\Phi)+\alpha\sqrt{g}F
 \sigma \ . \ee 
Where $\Phi$ is the $SO(g+1)$ multiplet
$\Phi=(\pi_1,\pi_2,...,\pi_g,\sigma)$ and the potential is
$V(\Phi)=-\mu^2\mid\Phi\mid^2+\lambda\mid\Phi\mid^4$. For
appropriate $\mu^2$ values the potential produces spontaneous
symmetry breaking from the $SO(g+1)$ symmetry down to $SO(g)$. As
in the NLSM we have added the last term to the Lagrangian in order
to break also explicitly the $SO(g+1)$ symmetry thus producing a
mass $m$ for the Goldstone bosons. Defining the vacuum as
$\Phi_{vac}=(0,0,...,0,\sigma_0)$ the LSM describes the dynamics
of $g$ pions of mass $m$ and a Higgs field $h=\sigma-\sigma_0$
with mass $m_h$  $(m_h^2=8\lambda g F^2+  3 m ^2)$. The Lagrangian
of the LSM in terms of these fields reads: \ba
\mathcal{L}=\frac{1}{2}\partial_{\mu}\pi^a\partial^{\mu}\pi^a-\frac{1}{2}m^2\pi^2+\frac{1}{2}\partial_{\mu}
h \partial^{\mu}h-\frac{1}{2} m_h^2 h^2  \nonumber  \\
 -\kappa h(\pi^2+h^2)-\lambda(\pi^2+h^2)^2, \ea
where $\kappa= 4\lambda\sqrt{gF^2+m^2/4\lambda}$. The NLSM
introduced in the previous section can be obtained  from this one
in the limit $m_h$ going to infinity thus showing that the NLSM
can be understood as the low energy effective theory of the LSM.

From this Lagrangian it is possible to obtain the elastic
scattering amplitude for pions perturbatively. However it is much
more interesting for our purposes here to consider another kind of
approach which is the large $g$ limit. The pion elastic scattering
amplitude for both the LSM and the NLSM for massive pions were
found in \cite{largeN}. One important thing concerning this
limit is that it is defined properly only if it is taken with $g
F^2$ fixed. Again we are interested in the non-relativistic limit.
Then the relevant low energy amplitude (Weinberg theorem) is in
this case $A=(s-m^2)/gF^2$ and therefore we have the effective
$f^2=g F^2$ fixed in the large $g$ limit. The averaged  cross
section in this limit is:
 \be \sigma_g=\frac{11 m^2}{128\pi (g^2F^4)} + O(1/g)  ,  \ee
the viscosity is: \be \eta_g = \frac{40\pi^{3/2}g^2F^4}{11
m^{3/2}}\sqrt{T} \ . \ee and finally we have: \be
\label{KSSlargeN} \frac{\eta_g}{s} =
\frac{80\sqrt{2}\pi^3}{11}\frac{(g^2F^4)}{m^4}
\frac{m}{T}\frac{1}{n\lambda^3\left( \log
\frac{g}{n\lambda^3}+\frac{5}{2}. \right)}\ . \ee This formula
provides the value of the KSS parameter in the non-relativistic
and classic regime of a $SO(g+1)/SO(g)$ NLSM in the large $g$
limit with $g F^2$ fixed. As this QFT descends from the
corresponding LSM we have found in principle an explicit example
of a non-relativistic system coming from a renormalizable QFT which
violates the KSS bound for large enough $g$.

Is there a way out for the KSS conjecture? In principle there
could be one. As we discussed above the NLSM incorporates processes
where two pions produce any even number of pions. Even if these
processes are very much suppressed at low temperatures $T<<m$,
they are always present. This in particular means that we cannot
fix the chemical potential at will and consequently an arbitrary
value of $\mu$ will correspond in general to a metastable state
outside chemical equilibrium. After some time this state
will relax to the absolute stable equilibrium state $\mu= m$. Thus
the density will  be no more arbitrary but it will be
completely determined by the temperature alone. Interestingly
enough if  $\mu = m$ then the condition $n<<g/\lambda^3$ cannot be
fulfilled and our computation here is not valid any more. 

One can think of introducing a flavor chemical potential (associated to 
the electric or strange charge), conserved by the strong interactions. 
These charges of course leak through the weak interactions to electrons 
and muons, and are lost to the pion gas. However, for a theoretical 
construction of a pure sigma model system, flavor is conserved. That is, 
although the total number of pions varies, the relative share between the 
different flavors can be maintained out of chemical equilibrium with the 
appropriate chemical potentials, and this is enough to violate the 
conjecture bound.

The conclusion is that the KSS violation that we have found here
applies only to states which are not chemically stable but not for
a genuine thermodynamic equilibrium state, unless one can force chemical 
potentials associated to flavor quantum numbers upon the system. 
It could be the case that the same reasoning applies to the KSS bound 
violations found in \cite{Cohen:2007qr} by using scaling arguments. This 
is because interaction terms changing the particle number probably appear 
in the Lagrangian density of any interacting relativistic quantum
field theory satisfying the Wightman axioms.

However there is no fundamental reason for the KSS conjecture
to hold in the strong sense, for example for non-relativistic systems of
complex molecules with exponentially large degeneracy $g$, which cannot
be described by the low energy limit of a relativistic QFT.
In the next section we introduce a class of systems that could
eventually provide a relevant physical example of KSS-bound violation.

\section{Isomeric molecules provide large degeneration factor}

There is no obvious classical gas that can be described by the
Non-Linear Sigma Model, because the spontaneous symmetry breaking
mechanism may not be active (Goldstone's theorem does not hold in
a non-relativistic quantum field theory).

However, to overcome the KSS bound by increasing the entropy, it
is sufficient to have a multicomponent gas with a large number of
species. The only reason we invoked the NLSM above was to write an
expression for the cross section at low energies that we could
control analytically in a model that can be UV completed. If we
are willing to take some uncertainty and accept an unknown
cross-section as a parameter instead of eq. (\ref{KSSchiral}),
 there should be no problem in decreasing the KSS number
below $1/{4\pi}$ provided we can arbitrarily increase the number
of components in the gas.

To obtain large number of particle species, monoatomic gases are out of
question as the count of stable isotopes is quite limited around the
stability valley. From the modest logarithmic growth of the mixing
entropy with the degeneracy exposed in the denominator of eq.
(\ref{KSSchiral}), and the typical values for a gas in figure
\ref{normalgases} and other works,  that are an order of magnitude above
the bound, we see that we need no less than 20 000 different species to
undercome the bound.

\begin{figure}
\psfig{figure=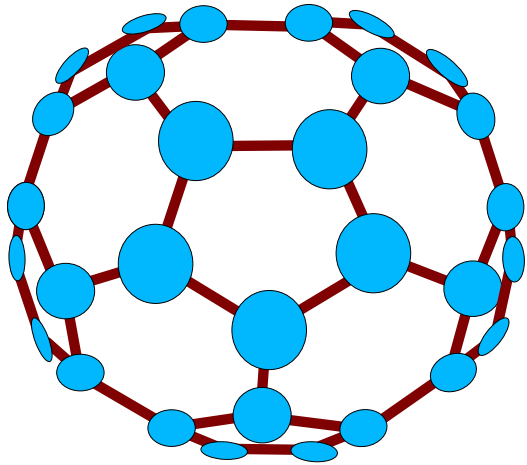,height=2.4in}
\caption{The Buckminsterfullerene, $C_{60}$. For temperatures
slightly about the sublimation point of about 750$K$,
the fullerene vapor provides a reasonable representation  of a hard-sphere
gas.
\label{fulereno1}}
\end{figure}

Molecular physics offers by far the largest variety of similar species in
terms of stable isomers. A popular molecule family that serves our purpose
for a {\emph{gedanken experiment}} is that of the fullerenes. We sketch in
figure \ref{fulereno1} the well-known Buckminsterfullerene (a $C_{60}$
truncated icosahedron).
Some two decades after their discovery, fullerenes are now copiously
produced (macroscopic fractions of a gram are usual) in the form of
powder, and have also been studied in disolution.
However, they are known
to sublimate to a fullerene vapor at a temperature of about 750 $K$.
 A review on
physical properties from where to track older literature is ref.
\cite{Gunnarson}.
The carbon atoms in the vertices of the truncated icosahedron $C_{60}$ all
fall at the surface of a sphere. Therefore, just above the sublimation
temperature, a dilute fullerene gas must behave as if a hard sphere gas,
with very suppressed rheologic and other properties that would blur
experimental data obtained with conventional hydrocarbon or polymer
chains.

But how does one obtain a large number of different species? There
is also a family of slightly more complicated molecules,
substituted fullerenes, where one or more Carbon atoms in the cage
surface are replaced, such as Boron-Nitrogen substituted
fullerenes (One can also talk of metallofullerenes, when a metal
atom such as Scandium or Lantanum is trapped {\emph{inside}} the
cage, but these are not our focuse now). Another system that could be 
useful is aggregated fullerenes, where one of the double carbon bonds 
may become simple and the extra carbon valence used to hook a 
submolecule. We refer to all simply as substituted fullerenes, 
without regard of whether the counting of sites refers to a given vertex 
or to a given side of the fullerene.

\begin{figure}
\psfig{figure=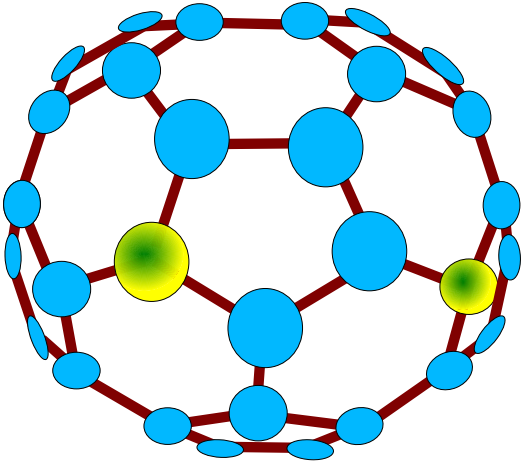,height=2.4in}
\caption{Example of doubly-substituted $C_{60}$.
\label{fulereno2}}
\end{figure}

Now, given enough substitutions, the power of combinatorial counting comes
to help. If, after accounting for the symmetries of a specific substituted
fullerene, there are $N$ possible sites (or edges) for the substitution,
and $M$ identical substituting atoms (perhaps not counting the first 
substitution), the number of different species in the gas sums up to
\be
g= \left(
\begin{tabular}{c}
N \\ M
\end{tabular}
\right)=\frac{N!}{M!(N-M)!}
\ee
that can be made quite large (for example, with $N=12$, $M=3$, one obtains
220 isomers).
In figure \ref{fulereno2} we plot an example $C_{60}$ with two
substitutions.
The substituted fullerenes are distinguishable particles (for example by
spectroscopic measurements of their vibrational lines) but have
essentially equal mass and very similar scattering cross section.
Their viscosity is then essentially independent of the number of different
isomers. However the entropy grows logarithmically with the number of
molecules and could eventually undercome the KSS bound. It seems however 
unlikely that the number of $C_{60}$ isomers will suffice \footnote{We 
thank D. Son for a rough estimate indicating this particular gas will 
still fail to violate the bound.}

But another variable comes into play: the density. The larger the 
molecular radius, the less dense that we can pack the molecules (and also 
the sooner the dilute gas approximation breaks down anyway). This makes 
the entropy density fall as $s\propto R^{-3}$ whereas the viscosity grows 
only as $\eta \propto R^2$. Therefore we need an estimate of what the 
number of species should be for $\eta/s =1/4\pi$. This happens, assuming a 
certain packing coefficient $c$ defined by $n=cR^-3$, for cubic packing
$c\sim 1/8$, when 
\be \label{neededg}
g=c \frac{\lambda^3}{R^3} e^{\frac{5}{2}\left(
\frac{\sqrt{\pi m T}R}{2c} -1 \right)} \ .
\ee
The inverse cubed radius multiplying the exponential cannot compete with 
the radius in the exponent, and the same applies to the $m$ and $T$ 
dependence. From this formula it is obvious that the number of species 
necessary needs to grow exponentially with their radius, mass and 
temperature. To minimize the required number of isomers, one therefore 
requires a gas that can stay cold without condensing, made of compact, 
light molecules. Helium satisfies all three conditions, but fails to 
violate the KSS bound by a factor 8.8, therefore we would need some 
$2400^{\frac{s}{n}\ar_{critical}}$ isomers!.

\begin{figure}
\psfig{figure=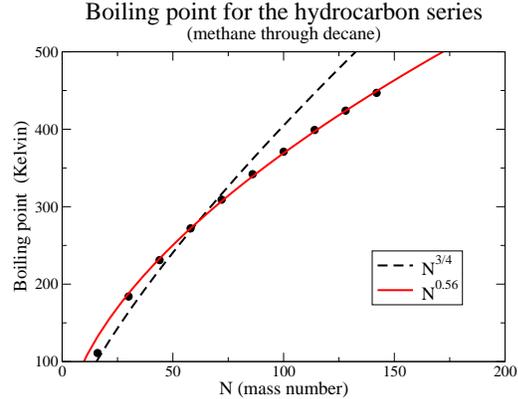,height=3.0in,angle=-90}
\caption{Boiling point for the hydrocarbon series. It can be fit by a 
function of the form $T_b(K)=28N^{0.56}$, growing with the molecule size 
because of the increased polarizability. \vspace{-0.3cm}
\label{boiling}}
\end{figure}

We can further reduce eq. (\ref{neededg}) by assuming that the mass of a 
molecule scales with the radius as  a power $m\propto R,\ R^2,\ R^3 $ for 
families of molecules such as polymers, fullerenes and globular proteines 
respectively. Further the boiling point is also known to scale as a power 
(typically smaller than 1) of the mass number, see fig. \ref{boiling} for 
an example.

Expressing everything in terms of the number of atoms in the molecule,
we find a typical
$$
g\propto N^{-7/2} e^{c N^{7/6}} 
$$
since the number of possible substitutions will typically grow as (in 
Stirling's approximation)
$$
g \propto N ! \simeq e^{N\log N} 
$$
we see that most families of substituted molecules will fail, since they 
will start further from the bound than Helium, but increasing their size 
provides barely, if at all, enough gain in isomer number.  
The candidate molecule family  needs 
to satisfy four conditions of compactness, coldness, lightness, and large 
isomer number. 
Helium satisfies the three first optimally, but fails to violate the 
bound by a factor about 8.8 at its critical point, and being chemically 
inert, it fails the fourth condition. 

\section{Discussion and outlook}

On physical grounds, it has been argued before \cite{Danielewicz:1984ww}
that the applicability of the Navier-Stokes equations restricts the
possible range of viscosities, and moreover, the viscosity should be
larger than the entropy density up to a constant factor, on the grounds
of the uncertainty principle alone. The novelty
\cite{KSS} is now that for a whole class of field theories, those with a
gravity dual, the factor is precisely $1/(4\pi)$.
No example is known with a smaller value.

From this note, it is apparent that a physically realizable
non-relativistic system can indeed evade this KSS bound. We have shown
this explicitly and as a matter of principle with the Non-Linear Sigma
Model, gave an example of multicomponent gas with hundreds 
to thousands of isomers, composed of substituted fullerenes, and given 
the conditions that can provide a physical violation of the bound.
In the end one can argue that this, or any other molecular gas made of
stable isotopes, may be completed by QED and QCD with the matter content
$e^-,u,d$ (and perhaps $s$). Since there are multiple such fluids
associated to the many scales and eigenvalues of the combined
Hamiltonian, the question arises on to which of them, or whether to all,
would a precisely stated KSS conjecture apply.

In spite of the results found here concerning the $\eta/s$ ratio at
the non-relativistic classic regime of the $SO(g+1)/SO(g)$ NLSM in
the large $g$ limit, it is possible that the KSS bound could be
maintained in the weak sense because of the particle production
present in any consistent QFT. However multicomponent gases of
complex carbon molecules as the ones presented here could violate
the bound in the strong sense.

It is finally interesting to speculate on a possible measurement of 
$\eta/s$ for the hadron gas in Relativistic Heavy Ion Collisions. The 
reason is that, if the traditional ideas of Hagedorn implying an 
exponential growth of the hadron species when approaching the phase 
transition, also built into the concept of Regge trajectories, would be 
correct, then $\eta/s$ would dramatically fall near the phase transition. 
A finite measurement of $\eta/s$ at RHIC, FAIR or the LHC, when combined 
with eq. (\ref{neededg}) or a suitable generalization thereof, could be 
used as an upper bound on the number of hadron resonances. To our 
knowledge this has never been attempted.

\emph{ 
We thank enthusiastic correspondence with T. Cohen and D. T. Son about 
possible improvements to the original preprint.
This work has been supported by
grants FPA 2004-02602, 2005-02327, PR27/05-13955-BSCH (Spain)}

\end{document}